\documentclass[12pt]{article}

\usepackage[numbers,sort&compress]{natbib}
\usepackage{graphicx}
\usepackage{amsmath}
\usepackage{amssymb}
\usepackage{hyperref}
\ifx\hypersetup\sadfkjashdfkxja\else
\hypersetup{pdftitle={Giant Gravitons on AdS4xCP3
and their Holographic Three-point Functions}}
\hypersetup{pdfauthor={S. Hirano, C. Kristjansen, D. Young}}
\hypersetup{pdfsubject={}}
\hypersetup{pdfkeywords={}}
\hypersetup{plainpages=false}
\hypersetup{bookmarksnumbered=true}
\hypersetup{pdfstartview=FitH}
\hypersetup{pdfpagemode=UseNone}
\hypersetup{colorlinks=false}
\hypersetup{citebordercolor={.5 1 .5}}
\hypersetup{urlbordercolor={.5 1 1}}
\hypersetup{linkbordercolor={1 .7 .7}}
\fi

\def\be{\begin{equation}}
\def\ee{\end{equation}}
\def\bsp{\be\begin{split}}
\def\la{\langle}
\def\ra{\rangle}

\def\G{\Gamma}
\def\D{\Delta}

\def\g{\gamma}

\def\d{\delta}
\def\e{\epsilon}
\def\m{\mu}

\def\n{\nu}

\def\l{\lambda}

\def\vt{\vartheta}

\def\bC {\mathbb{C}}

\def\bZ {\mathbb{Z}}

\def\vp{\varphi}

\makeatletter

\newcommand{\Rmnum}[1]{\expandafter\@slowromancap\romannumeral #1@}
\makeatother

\newcommand{\beq}{\begin{equation}}
\newcommand{\eeq}{\end{equation}}
\newcommand{\bea}{\begin{eqnarray}}
\newcommand{\eea}{\end{eqnarray}}

\renewcommand{\title}[1]{\vbox{\center\LARGE{#1}}\vspace{5mm}}
\renewcommand{\author}[1]{\vbox{\center\large{#1}}\vspace{5mm}}
\newcommand{\address}[1]{\vbox{\center\em#1}}

\newcommand{\Tr}{\mathrm{Tr}}

\newcommand{\Cset}{\bC}

\newcommand{\comment}[1]{}
\begin{document}
\bibliographystyle{utphys}
\newpage
\setcounter{page}{1}
\pagenumbering{arabic}
\renewcommand{\thefootnote}{\arabic{footnote}}
\setcounter{footnote}{0}

\begin{titlepage}
\title{\vspace{1.0in} {\bf Giant Gravitons on $AdS_4\times \Cset \mathrm{P}^3$
and their Holographic Three-point Functions}}
 
\author{S.\ Hirano$^a$, C.\ Kristjansen$^b$, D.\ Young$^b$}

\address{$^a$Department of Physics, Nagoya University, Nagoya 464-8602, Japan\\
$^b$Niels Bohr Institute, Blegdamsvej 17, DK-2100 Copenhagen, Denmark}

%\email{$^a$ $^b$kristjan, dyoung@nbi.dk}

\abstract{We find a simple parametrization of the anti-symmetric giant
  graviton in $AdS_4\times\Cset \mathrm{P}^3$, first constructed
  in~\cite{Giovannoni:2011pn}, dual to the anti-symmetric Schur
  polynomial involving two bi-fundamental complex scalar fields of
  ABJM theory. Using this parametrization we evaluate in a
  semi-classical approach the three-point function of two such giant
  gravitons and one point-like graviton considering both extremal and
  non-extremal configurations. We likewise discuss the case of the
  symmetric giant graviton in $AdS_4\times \Cset \mathrm{P}^3$. 
  Finally, we
  provide an expression for the planar three-point function of chiral primary
  operators in ABJM at strong coupling  and find
  that the results for the giant graviton three-point functions reduce
  to this expression in the point-like limit.}

\end{titlepage}

%%%%%%%%%%%%%%%%%%%%%%%%%%%%%%%%%%%%%%%%%%%%%%%%%%%%%%%%%%%%%%%%%%%%%%%
\section{Introduction}
\label{Introduction}
Giant gravitons constitute an important entry in the AdS/CFT
dictionary. In the gravity language giant gravitons represent
extended higher dimensional objects, D- or M-branes, while in the
field theory language they correspond to operators carrying higher
representations of the gauge group. In particular, the latter
characterization imply that giant gravitons encode information about
finite-$N$ gauge theory.

In the $AdS_5\times S^5$ case~\cite{Maldacena:1997re} one has a simple
and beautiful description of 1/2 BPS giant gravitons. On the string
theory side the giant gravitons are D3-branes which wrap an $S^3$
inside either $S^5$ or $AdS_5$ while moving on a circle of $S^5$
with a fixed angular momentum~\cite{McGreevy:2000cw,Grisaru:2000zn,
  Hashimoto:2000zp,Balasubramanian:2001nh}. The gauge theory dual of
the $S^5$ giant graviton is the completely anti-symmetric Schur
polynomial of a single complex scalar while the dual of the $AdS_5$
giant is the completely symmetric Schur
polynomial~\cite{Balasubramanian:2001nh,Corley:2001zk}.

For the $AdS_4\times \Cset\mathrm{P}^3$ case~\cite{Aharony:2008ug} the
situation is slightly more complicated. The simplest possible Schur
polynomials are constructed out of two complex bi-fundamental scalars,
see~\cite{Dey:2011ea,Chakrabortty:2011fd} for a discussion of these.
The gravity dual of the completely symmetric Schur polynomial is a
D2-brane which wraps an $S^2$ inside $AdS_4$ and (after uplift to
M-theory) rotates along a great circle of $S^7$ orthogonal to the
compactification
circle~\cite{Berenstein:2008dc,Nishioka:2008ib,Hamilton:2009iv}. The
gravity dual of the anti-symmetric Schur polynomial can be described
in M-theory language as an M5-brane which wraps two $S^5$'s
intersecting at an $S^3$, all inside $S^7$, and which like its
symmetric cousin rotates along a circle orthogonal to the
compactification circle. Its maximal version was discussed
in~\cite{Berenstein:2008dc,Murugan:2011zd}, see
also~\cite{Gutierrez:2010bb, Lozano:2011dd,Herrero:2011bk}, but the
general solution was first constructed in~\cite{Giovannoni:2011pn}.

In the present letter we find an improved parametrization of the
anti-symmetric giant graviton of $AdS_4\times \Cset\mathrm{P}^3$. This
greatly simplified parametrization allows us to calculate analytically
the three-point function involving two such giant gravitons and one
point-like graviton in the holographic approach suggested for strings
in~\cite{Zarembo:2010rr,Costa:2010rz,Janik:2010gc,
  Buchbinder:2010ek,Buchbinder:2010vw} and generalized to branes
in~\cite{Bak:2011yy, Bissi:2011dc}. Unlike what is the case for ${\cal
  N}=4$ SYM, in ABJM theory three-point functions of 1/2 BPS chiral
primary operators are not protected. The chiral primary operators are
built from pairs of the four bi-fundamental complex scalar fields $W_I$ of the
ABJM theory and their conjugates $\bar W^I$ and are given by
\be\label{defop}
{\cal O}_{A} = \frac{(4\pi)^{J/2}}{\sqrt{J/2}\,\lambda^{J/2}}\,
({\cal C}_{A})^{I_1\ldots I_{J/2}}_{K_1\ldots K_{J/2}} \,
\Tr \left(W_{I_1} \bar W^{K_1} \cdots W_{I_{J/2}} \bar W^{K_{J/2}}\right),
\ee
where $\lambda = N/k$ is the 't Hooft coupling of ABJM, and ${\cal C}_{A}$ is
completely symmetric in upper and (independently) in lower indices,
while the trace taken on any pair consisting of one 
upper and one lower index
vanishes. The tensors are orthonormal, so that
\be
({\cal C}_{A})^{I_1\ldots I_{J/2}}_{K_1\ldots K_{J/2}} 
(\bar{\cal C}_{B})^{K_1\ldots K_{J/2}}_{I_1\ldots I_{J/2}} =\delta_{AB} ,
\ee
and the two-point function is protected and given by $\la {\cal
  O}_{A}(x) \bar{\cal O}_{B}(0) \ra = \delta_{AB}/|x|^{J}$. The
three-point function structure constants $C_{123}$ are then defined as
\be
\la {\cal O}_1(x_1) {\cal O}_2(x_2) {\cal O}_3(x_3) \ra = \frac{C_{123}}
{|x_1 - x_2|^{\g_3} |x_2 - x_3|^{\g_1} |x_3 - x_1|^{\g_2}},
\ee
where $\gamma_i = (\sum_jJ_j-2J_i)/2$. Using the fact that the chiral
primary operators are dual to point-like gravitons, the structure
constants were calculated at strong coupling and large-$N$ long ago
\cite{Bastianelli:1999en} using M-theory on $AdS_4\times S^7$ and so
corresponding to the case of ABJM with Chern-Simons level $k=1$. This
expression may then be generalized for arbitrary $k$, see appendix
\ref{app:cpo}. The result is (we take $J_3\geq J_2\geq J_1$)
\bsp\label{abjmcpo}
&C_{123}^{\l\gg 1} =\\
&\frac{1}{N} \left(\frac{\l}{2\pi^2}\right)^{1/4}
\frac{\prod_{i=1}^3 
\sqrt{J_i+1}\,(J_i/2)!\,\G(\g_i/2+1) }{\Gamma(\g/2+1)} \\
&\sum_{p=0}^{\g_3}
\frac{\left({\cal C}_1\right)^{I_1\ldots I_p I_{p+1}\ldots I_{J_1/2}}
_{K_1 \ldots K_{\g_3 -p} K_{\g_3-p+1} \ldots K_{J_1/2}}
\left({\cal C}_2\right)^{K_1 \ldots K_{\g_3 -p} L_1\ldots
L_{\g_1-J_2/2+p}}_{I_1\ldots I_p M_1 \ldots M_{J_2/2-p}}
\left({\cal C}_3\right)^{K_{\g_3-p+1} \ldots K_{J_1/2}M_1 \ldots
  M_{J_2/2-p}}
_{I_{p+1}\ldots I_{J_1/2}  L_1\ldots L_{\g_1-J_2/2+p}}}
{p! (\g_3-p)! (\g_1-J_2/2+p)!
(J_2/2-p)!(\g_2-J_1/2+p)! (J_1/2-p)!}
.
\end{split}
\ee 
where $\gamma = \gamma_1+\gamma_2+\gamma_3$. In contrast to the ${\cal
  N}=4$ SYM case, we see that not only is there a $\l$-dependence, but
also that there is a range of contractions of the ${\cal C}$
tensors. This freedom amounts to the number $p$ of upper indices in
${\cal C}_1$ which are contracted with lower indices in ${\cal
  C}_2$. It is instructive to compare this result to the tree-level
perturbative result, where to the leading order in $1/N$, only one
such contraction can appear\footnote{The range of contractions in
  (\ref{abjmcpo}) includes all possible contractions, and so naturally
  includes the planar contraction.}, which we denote by $\la {\cal
  C}_1\,{\cal C}_2\,{\cal C}_3\ra_{\text{planar}}$. One then has
\be
C_{123}^{\l\ll 1} = \frac{1}{N}
\sqrt{(J_1/2)(J_2/2)(J_3/2)}\, \la {\cal C}_1\,{\cal
  C}_2\,{\cal C}_3\ra_{\text{planar}} + {\cal O}(\l^2/N).
\ee
Therefore we see that the chiral primary structure constant $C_{123}$
is a highly non-trivial function of both the coupling $\lambda$ and
the charges defining the operators. In the extremal case, when $J_3 =
J_1+J_2$, the result at strong coupling simplifies dramatically, and
only the planar contraction remains. One finds
\be
C_{123}^{\l\gg 1} |_{J_3 = J_1 + J_2}=
\frac{1}{N} \left(\frac{\l}{2\pi^2}\right)^{1/4}
\sqrt{(J_1+1)(J_2+1)(J_3+1)}\, \la {\cal C}_{J_1}{\cal
  C}_{J_2}{\cal C}_{J_3}\ra_{\text{planar}}.
\ee
It is a very interesting direction of future research to determine the
$C_{123}$ at higher (or at all) orders of perturbation theory. Judging
from the similarity between the strong coupling and tree-level results
in the extremal case, it would appear that the extremal problem is far
more tractable.

In this paper we provide a generalization of (\ref{abjmcpo}) to a
specific case when two of the 1/2 BPS operators correspond to a
specific giant graviton, and the remaining operator to a pointlike
graviton. This implies taking two of the charges, $J_2$ and $J_3$,
large and the remaining one, $J_1$, to be order one. We will find that
taking the large $J_2=J_3$ limit of (\ref{abjmcpo}) coincides with the
small $J_2/N=J_3/N$ limit of the expressions we obtain for the
two-giant, one point-like three-point structure constants.\footnote{In
  the latter limit we first take $N, J_i\to\infty$ with $J_i/N$ fixed
  and then $J_i/N$ small (where $i=2,3$).} This behaviour was also
observed in the context of ${\cal N}=4$ SYM \cite{Bissi:2011dc} where
similar holographic three-point functions involving giant gravitons
were studied. There, in addition, it was found that extremal
correlators exhibited a structure very similar to the dual gauge
theory correlators at tree level whereas a complete match was not
observed. Later it was shown that one does obtain a complete match for
non-extremal three-point functions~\cite{Caputa:2012yj}. In the
present letter we present three-point functions of both types
expecting that the non-extremal ones truly reflect the strong coupling
behaviour of ABJM theory and hoping that the others could help shed
light on the subtleties of the extremal case. We also calculate the
three-point functions in ABJM perturbation theory at
tree-level. Unsurprisingly, it is clear that like in the point-like
case, a non-trivial function of the coupling and the charges defining
the operators interpolates between weak and strong coupling.

We start by introducing the coordinate system which naturally leads to
our improved parametrization in section~\ref{coordinate} and move on
to discussing in detail the anti-symmetric giant in
section~\ref{antisymmetric}. In particular, we calculate in this
section a number of extremal as well as non-extremal three-point
functions involving two anti-symmetric giants and one point-like
graviton. The same type of correlation functions are subsequently
computed for symmetric giant gravitons in section~\ref{symmetric} and
in appendix \ref{appendix} the dual ABJM three-point functions at
tree-level are calculated. Finally, section~\ref{conclusion} contains
our conclusions.

%%%%%%%%%%%%%%%%%%%%%%%%%%%%%%%%%%%%%%%%%%%%%%%%%%%%%%%%%%%%%%%%%%%%%%%
\section{The coordinate system \label{coordinate}}

For the study of the anti-symmetric giant graviton in $AdS_4\times
\Cset\mathrm{P}^3$, it will prove particularly convenient to use the
parametrization
\begin{equation}
Z_1 = r e^{i\left(\chi/2+\phi\right)}Z\ ,\qquad
Z_2 = r e^{-i\left(\chi/2-\phi\right)}\bar{Z}^{-1}\ ,\qquad
Z_3 = e^{\rho_3+i(\theta_3+\phi)}\ ,\qquad
Z_4=r_4e^{i\phi}\ ,\label{coordinates}
\end{equation}
where $r_4^2=1-2r^2\cosh(2\rho)-e^{2\rho_3}$ and
$Z=e^{\rho+i\theta}$. The $Z_I$ cover the unit $S^7$
\begin{equation}
|Z_1|^2+|Z_2|^2+|Z_3|^2+|Z_4|^2=1\ ,\label{S7}
\end{equation}
once for 
\bsp
&0\le e^{2\rho_3}\le 1-2r^2\cosh(2\rho)\equiv
e^{2\rho_3^{\rm max}}, \\
-&\rho_{\rm max} \leq \rho \leq \rho_{\rm
  max} ~~\text{where}~~ \cosh(2\rho_{\rm max})=1/(2r^2),\\ 
&0\leq r\leq1/\sqrt{2}, \quad 
0\le\theta,\,
\theta_3,\, \chi,\, \phi \le 2\pi.
\end{split}
\ee
The $Z_I$ are also in one-to-one correspondence with the four
bi-fundamental complex scalars $W_I$ of the ABJM theory. In terms of
$z_i\equiv Z_i/Z_4$ ($i=1,2,3$), the $S^7$ metric is expressed as the
$U(1)$ fibration over the Fubini-Study $\Cset\mathrm{P}^3$,
\begin{equation}
ds_{S^7}^2={dz_id\bar{z}_j\over (1+z_k\bar{z}_k)^2}
\left[\delta_{ij}(1+z_k\bar{z}_k)-\bar{z}_iz_j\right]
+\left(d\phi+A\right)^2\ ,
\end{equation}
with the standard 1-form $A={i\over 2}
(d-\bar{d})\ln(1+z_k\bar{z}_k)$. The angle $\phi$ is the coordinate
parametrizing the M-theory circle. The $S^7/\bZ_k$ is obtained by
restricting the range of the angle $\phi$ to $0\le \phi\le 2\pi/k$.

%%%%%%%%%%%%%%%%%%%%%%%%%%%%%%%%%%%%%%%%%%%%%%%%%%%%%%%%%%%%%%%%%%%%%%%%%%%
\section{The anti-symmetric giant graviton \label{antisymmetric}}

The giant graviton dual to the Schur polynomial of the $U(N)$ adjoint
field $W_1\bar{W}^2$ in antisymmetric representations was found
in~\cite{Giovannoni:2011pn}. In M-theory, the giant graviton is an
M5-brane in $S^7/\bZ_k$ and described by the curve
\begin{equation}
Z_1\bar{Z}_2 = \alpha^2 e^{i\chi(t)}\ ,
\end{equation}
where $\alpha$ is the constant related to the size of the giant. The
time $t$ is that of the global AdS space, and the giant graviton
rotates in the $\chi$-direction. This curve reflects the property that
the Schur polynomial of the maximal dimension $N$ becomes a product of
di-baryon operators, $\det W_1\det\bar{W}^2$. Namely, when the giant
is maximal ($\alpha=0$), the curve becomes two $S^5$'s ($Z_1=0$ and
$Z_2=0$) intersecting at an $S^3$ ($Z_1=Z_2=0$) dual to a product of
two di-baryons.

In terms of the coordinates introduced in the previous section, the
world volume of the M5 giant is spanned by $(t, \rho, \rho_3, \theta,
\theta_3, \phi)$ where $-\rho_{\rm max} \leq \rho \leq \rho_{\rm
  max}$ where $\cosh(2\rho_{\rm max})=1/(2\alpha^2)$, $0\le e^{2\rho_3}\le
1-2\alpha^2\cosh(2\rho)\equiv e^{2\rho_3^{\rm max}}, 0\le\theta,
\theta_3, k\phi \le 2\pi$.

%-------------------------------------------------------------------------%
\subsection{The probe analysis}

We shall work in the probe approximation. It is straightforward to
find the low energy effective action, i.e., the DBI $+$ WZ
action, for the M5 giant:
\bsp
&S_{\rm DBI}=\\
&-{(2\pi)^3\over k}T_{\rm M5}R_{\rm
  S^7}^6\alpha^2\!\!\int_{-\infty}^{\infty} {dt\over 2}
\int_{-\rho_{\rm max}}^{\rho_{\rm max}} \!\!\!\!\!\! d\rho \int_{0}^{e^{2\rho_3^{\rm
      max}}} \!\!\!\!\!\! \!\!\! de^{2\rho_3}
\sqrt{\left(\cosh(2\rho)-2\alpha^2\omega^2\right)
  \left(\cosh(2\rho)-2\alpha^2\right)}\ ,
\end{split}
\ee
where $T_{\rm M5}=\ell_P^{-6}/(2\pi)^5$, $R_{S^7}^6=(2R_{\rm
  AdS})^6=2^3(2\pi)^2kN\ell_P^6$, and $\omega\equiv {d\chi\over dt}$.
Meanwhile, the background 6-form potential is given by
\begin{equation}
C_6= 2T_{\rm M5}R_{S^7}^6e^{2\rho_3}r^2\left(r^2-{1\over 2}\cosh(2\rho)\right)d\rho\wedge d\rho_3\wedge d\theta\wedge d\chi\wedge d\theta_3\wedge d\phi+\cdots\ ,
\end{equation}
with an appropriate gauge choice.\footnote{The 6-form potential is
  proportional to $(A+d\Lambda)\wedge dA\wedge dA\wedge d\phi$ where
  the 1-form
  $A=2r^2\cosh(2\rho)d\theta+2r^2\sinh(2\rho)d\chi+e^{2\rho_3}d\theta_3$. This
  gauge corresponds to the choice $\Lambda=-\theta$.} We then find
that
\begin{equation}
S_{\rm WZ}=+{(2\pi)^3\over k}T_{\rm
  M5}R_{S^7}^6\alpha^2\int_{-\infty}^{\infty} {dt\over 2}
\int_{-\rho_{\rm max}}^{\rho_{\rm max}}\!\!\!\!\!d\rho \int_{0}^{e^{2\rho_3^{\rm
      max}}}
\!\!\!\!\!\!\!de^{2\rho_3}\left(\cosh(2\rho)-2\alpha^2\right)\omega\ .
\end{equation}
Introducing the new variable $x=\cosh(2\rho)$, the DBI $+$ WZ action for the M5 giant yields
\begin{equation}
S_{\rm M5}=8N\alpha^4\!\!\int_{-\infty}^{\infty} dt \int_{1}^{1\over
  2\alpha^2} \!\!\!dx {\left(x-{1\over 2\alpha^2}\right)
  \over\sqrt{x^2-1}}\left[\sqrt{\left(x-2\alpha^2\omega^2\right)
    \left(x-2\alpha^2\right)}-\omega
  \left(x-2\alpha^2\right)\right]\ .\label{M5giant}
\end{equation}
Note that the action vanishes when $\omega=1$ which corresponds to the
M5 giant moving at the speed of light, i.e., the giant graviton
solution.

The R-charge/angular momentum of the M5 giant is fixed
\begin{equation}
L\equiv{\partial L_{\rm M5}\over\partial
  \omega}=-8N\alpha^4\int_{1}^{1\over 2\alpha^2} dx {\left(x-{1\over
    2\alpha^2}\right)
  \over\sqrt{x^2-1}}\left[2\alpha^2\omega\sqrt{x-2\alpha^2\over
    x-2\alpha^2\omega^2} +\left(x-2\alpha^2\right)\right]\ ,
\end{equation}
 where $L_{\rm M5}$ is the Lagrangian for the M5 giant. As
 in~\cite{Giovannoni:2011pn}, we are unable to find $\omega$ or the
 Routhian $R(\alpha, L)\equiv L\omega-L_{\rm M5}(\alpha, \omega)$ as a
 function of $L$ and $\alpha$. However, we know that the Routhian is
 minimized when $\omega=1$, as numerically checked
 in~\cite{Giovannoni:2011pn}. Hence the energy $E$ of the giant
 graviton is equal to $L$, saturating the BPS bound. This agrees with
 the scaling dimension of the Schur polynomial, as $L$ counts
 field-pairs, i.e. $W_1\bar W^2$ which have conformal dimension $1$.

It is easy to find the relation between the angular momentum $L$ and
the parameter $\alpha$ related to the size of the giant graviton
($\omega=1$):
\begin{equation}
{L\over N}=\sqrt{1-4\alpha^4}
-4\alpha^4\log{1+\sqrt{1-4\alpha^4}\over 2\alpha^2}\ .
\end{equation}
The size is maximal when $\alpha=0$ and zero when
$\alpha={1\over\sqrt{2}}$. In the former case, the angular momentum is
maximal $L=N$ (stringy exclusion principle), whereas it vanishes,
$L=0$, in the latter case.

The dimensional reduction to type IIA is straightforward. The M5 giant
becomes a D4-brane. In particular, the maximal giant wraps two
$\Cset\mathrm{P}^2$'s intersecting at a $\Cset\mathrm{P}^1$.

%%%%%%%%%%%%%%%%%%%%%%%%%%%%%%%%%%%%%%%%%%%%%%%%%%%%%%%%%%%%%%%%%%%%%%%%%%%
\subsection{Holographic three-point functions}
\label{antiholo}

The three-point function between two of the giant gravitons described
in the previous section and a chiral primary operator, corresponding
to a point-like graviton, may be computed using the techniques
described in \cite{Zarembo:2010rr,Costa:2010rz,Bissi:2011dc}. The supergravity
fluctuations corresponding to a chiral primary operator have been
derived in \cite{Biran:1983iy,Castellani:1984vv,Bastianelli:1999bm},
and used in a very similar context to the present one in
\cite{Drukker:2008jm}, to which we refer the reader for details. The
fluctuations are given by
\begin{equation}\label{fluct}
\begin{split}
&\delta g_{\m \n} = h_{\m\n}=
\frac{4}{J+2}
\left[\nabla_\m\nabla_\n -\frac{1}{6} J(J-1)\right] \,s^J(X)\,Y_J(\Omega), \\
&\delta g_{\alpha \beta } = \frac{J}{3}\,g_{\alpha \beta
} \,s^J(X)\,Y_J(\Omega),\\
&\delta
C_{\mu_1\mu_2\mu_3}=2\,\e_{\mu_1\mu_2\mu_3\mu_4}\nabla^{\mu_4}
\,s^J(X)\,Y_J(\Omega),\\
&\delta C_{\alpha_1\ldots\alpha_6}=-2\,\e_{\alpha_1\ldots\alpha_7}\nabla^{\alpha_7}
\,s^J(X)\,Y_J(\Omega),\\
\end{split}
\end{equation}
where early greek indices refer to $S^7/\bZ_k$ coordinates $\Omega$
while late greek indices refer to $AdS_4$ coordinates $X$, $g$ is the
metric and $C_3$ and $C_6$ are the three-form and six-form
Ramond-Ramond potentials. $Y_J(\Omega)$ represents a scalar spherical
harmonic\footnote{\label{sphnrm}We take the radius of the $S^7/\bZ_k$
  to be 2, and the normalization of the spherical harmonics is given
  by $\int_{S^7/\bZ_k} Y^J (Y^K)^* =
  \delta^{JK}2^8\pi^4k^{-1}[(J/2)!]^2/(J+3)!$.} on $S^7/\bZ_k$ with
angular momentum $J$, while $s^J(X)$ is a scalar field propagating on
$AdS_4$ with mass-squared $J(J-6)/4$ and has a bulk-to-boundary
propagator given by
\be\label{b2b}
\la s^J(x,z) \, s^J(x_B,0)\ra=
\ell_p^{9/2}\frac{2^{J/2-1}\pi \sqrt{k}}{R_{\rm AdS}^{9/2}} \frac{J+2}{J}\sqrt{J+1} 
\frac{z^{J/2}}{((x-x_B)^2+z^2)^{J/2}} ,
\ee
where $(x,z)$ are Poincar\'e coordinates on $AdS_4$, and $x_B$
represents the boundary point. In terms of global coordinates $ds_{\rm
  AdS}^2=R_{\rm AdS}^2\left(-\cosh^2\mu dt^2+d\mu^2+\sinh^2\mu
d\Omega_2^2\right)$
\be\label{poin}
(x^0,x^1,x^2,z) = \frac{{\cal R}}{2(\cosh\m\cos t -n_1\sinh\m)}
(\cosh\m\sin t,\,n_2\sinh\m,\,
n_3\sinh\m,\,1),
\ee
where $\vec n\cdot\vec n=1$ and ${\cal R}$ denotes the separation
along the boundary of the two giant gravitons. The fluctuation
calculation is carried out in Euclidean space, so that $t\to -i t$ and $x^0 \to
-i x^0$. The fluctuations (\ref{fluct}) are pulled-back onto the
Euclidean M5-brane while the
field $s^J(X)$ is replaced by its bulk-to-boundary propagator
connecting the insertion point on the brane to the boundary. The
boundary point $x_B$, representing the position of the chiral primary
field, is sent to infinity and the resulting expression is integrated
over the on-shell world volume of the brane. The result is $-({\cal
  R}/(2 x_B^2))^{J/2}$ times the structure constant defining the
three-point function. Note that for the antisymmetric giant of the
previous section $\m=0$.

We find the following fluctuations of the DBI and WZ parts of the Euclidean
M5-brane Lagrangian density ($S=\int d^6\sigma {\cal L}$)
\bsp\label{this}
&\delta {\cal L}_{\rm DBI} = \frac{R_{\rm AdS}^6}{(2\pi)^5\ell_p^6}\,Y_J(\Omega) \frac{1}{2} 
\sqrt{g} \left( 2J +\frac{\cosh 2\rho}{\cosh
  2\rho - 2\alpha^2\omega^2}
\left[\frac{4}{J+2}\partial_t^2-\frac{J^2}{J+2}\right]\right)\,s^J(X),\\
&\delta {\cal L}_{\rm WZ} =  -\frac{R_{\rm AdS}^6}{(2\pi)^5\ell_p^6} \,\omega\,
\sqrt{g_{S^7/\bZ_k}}\, g_{S^7/\bZ_k}^{r\beta} \partial_\beta
Y_J(\Omega) \Biggl|_{r=\alpha}\, s^J(X),
\end{split}
\ee
where $\sqrt{g} = 2^6 \alpha^2 e^{2\rho_3} \sqrt{\left( \cosh
  2\rho-2\alpha^2\right) \left( \cosh 2\rho -2 \alpha^2
  \omega^2\right)} $, and $g_{S^7/\bZ_k}$ is the metric on the
$S^7/\bZ_k$ of radius 2, parametrized as in section \ref{coordinate},
so that on the classical solution, the coordinate $r$ is set to
$\alpha$. Using the following spherical harmonic, dual to the ABJM
theory operator $\Tr (W_1\bar{W}^2)^{J/2}$, we get the extremal
correlator with a point-like graviton which is the degeneration of the
antisymmetric giant graviton to a point
\be\label{point}
Y_J(\Omega) = \left( r^2 e^{i\chi} \right)^{J/2}.
\ee
The resulting structure constant is
\be\label{Canti}
C_{L,L-\Delta,\Delta}^A = 
\frac{1}{N}\left(\frac{\lambda}{2\pi^2}\right)^{1/4}
 2L \,\left(2\alpha^2\right)^\Delta\, \sqrt{2\Delta+1},
\ee
where $\lambda=N/k$, $L$ is the number of field-pairs (i.e. $W_1\bar W^2$) in the
giant graviton and $\Delta=J/2$ is the number of field-pairs in the chiral
primary.

Given the subtleties associated with extremal correlators appearing in
analogous computations in the $AdS_5/CFT_4$ correspondence
\cite{Bissi:2011dc}, where agreement between gauge theory and
holographic three point functions involving two giant gravitons is
found only for the non-extremal case~\cite{Caputa:2012yj}, we present here some
calculations of non-extremal correlators. Specifically we consider the
following operators \cite{Drukker:2008jm}
\be
\begin{aligned}
{\cal O}_{1,0}&=\frac{2\pi}{\sqrt{3}\lambda}
\Tr\Big[W_I\bar W^I-4W_1\bar W^1\Big],\\
{\cal O}_{2,0}&=\frac{8\pi^2}{3\sqrt{5}\lambda^2}
\Tr\Big[(W_I\bar W^I)^2-10W_I\bar W^I\,W_1\bar W^1+15(W_1\bar W^1)^2\Big],\\
{\cal O}_{3,0}&=\frac{16\pi^3}{3\sqrt{105}\lambda^3}
\Tr\Big[(W_I\bar W^I)^3-18(W_I\bar W^I)^2\,(W_1\bar W^1)\\
&\qquad\qquad\qquad\quad
+63(W_I\bar W^I)\,(W_1\bar W^1)^2-56(W_1\bar W^1)^3\Big],
\end{aligned}
\label{opss}
\ee
which are normalized chiral primary operators. The
corresponding spherical harmonics $Y_J(\Omega)$ are found by substituting
(\ref{coordinates}) for the $W_I$ and normalizing according to
footnote \ref{sphnrm}
\bsp 
&Y_2 = \frac{1}{2\sqrt{3}} \left( 1- 4\, r^2
e^{2\rho}\right),\\ 
&Y_4 = \frac{1}{3\sqrt{10}} \left( 1 - 10\, r^2
e^{2\rho}+15\, r^4 e^{4\rho}\right),\\ 
&Y_6 = \frac{1}{4\sqrt{35}}
\left(1 -18\, r^2 e^{2\rho} +63\, r^4 e^{4\rho} -56\, r^6 e^{6\rho}
\right).
\end{split}
\ee 
The calculation then proceeds similarly to the extremal case. Using
(\ref{this}) one obtains 
\be\label{aslo}
C^A_{{\cal O}_{\Delta,0}} 
= \frac{1}{N} \left( \frac{\lambda}{2\pi^2}\right)^{1/4} 
\times
\begin{cases}
-\frac{\pi}{2} L,\qquad &\Delta=1\\
-2\sqrt{2} L + \frac{11\sqrt{2}N}{9} (1-4\alpha^4)^{3/2},\qquad &\Delta=2\\
-\frac{9\pi}{32\sqrt{5}} \left(2 L (4+3\alpha^4) 
-7N(1-4\alpha^4)^{3/2}\right),\qquad &\Delta=3
\end{cases},
\ee
for the structure constant corresponding to the three-point functions
involving two giant gravitons and one of ${\cal O}_{1,0}$, ${\cal
  O}_{2,0}$, or ${\cal O}_{3,0}$ respectively. As mentioned in the
introduction, taking the small $L/N$ (i.e. $\alpha \to 1/\sqrt{2}$)
limit, in both the extremal and non-extremal cases, one obtains
agreement with the large $J_2=J_3=2L$ limit of the point-like result
(\ref{abjmcpo}), as was the case in ${\cal N}=4$ SYM
\cite{Bissi:2011dc}. Comparing with free-field contractions
(\ref{exttree}) and (\ref{nonxgg})-(\ref{nonxgg1}), we see that, as in
the point-like case, a non-trivial function of $\lambda$ and the
charges interpolates between weak and strong coupling results.

%%%%%%%%%%%%%%%%%%%%%%%%%%%%%%%%%%%%%%%%%%%%%%%%%%%%%%%%%%%%%%%%%%%
\section{The symmetric giant graviton\label{symmetric}}

The giant graviton dual to the Schur polynomial of the $U(N)$ adjoint
field $W_1\bar{W}^2$ in symmetric representations is an M2-brane
wrapping the $S^2$ in global AdS$_4$ space and rotating along the
great circle of $S^7/\bZ_k$. This is the so-called AdS giant
\cite{Grisaru:2000zn, Hashimoto:2000zp}. More specifically, the AdS
giant of interest to us rotates along the great $\chi$-circle;
$r={1\over\sqrt{2}}, \rho=e^{\rho_3}=\theta=\phi=0$, and
$Z_1\bar{Z}_2={1\over 2}e^{i\chi(t)}$.

The DBI $+$ WZ action for the M2 giant is given by
\begin{equation}\label{symmS}
S_{\rm M2}=-4\pi T_{\rm M2} R_{\rm AdS}^3\int
dt\left[\sinh^2\mu\sqrt{\cosh^2\mu-\omega^2}-\sinh^3\mu\right]\ ,
\end{equation}
where $ds_{\rm AdS}^2=R_{\rm AdS}^2\left(-\cosh^2\mu
dt^2+d\mu^2+\sinh^2\mu d\Omega_2^2\right)$ and $\omega\equiv
{d\chi\over dt}$. Note that $4\pi T_{\rm M2} R_{\rm
  AdS}^3=N/\sqrt{2\lambda}$. The angular momentum yields
\begin{equation}
L\equiv{\partial L_{\rm M2}\over\partial\omega}={N\over
  \sqrt{2\lambda}}{\omega
  \sinh^2\mu\over\sqrt{\cosh^2\mu-\omega^2}}\ .
\end{equation}
The Routhian $R(\mu, L)=L\omega-L_{\rm M2}(\mu, \omega)$ is minimized
at $\omega=1$ corresponding to the M2 giant moving at the speed of
light. The energy $E$ of the giant graviton is again $L$, saturating
the BPS bound and the size of the giant is related to the angular
momentum by
\begin{equation}
\sinh\mu = \sqrt{2\lambda}{L\over N}\ .
\end{equation}
There is no upper bound on the size of the giant in this case. The
dimensional reduction to type IIA is trivial, and the M2 giant becomes
a D2-brane.

A comment is in order: There is another type of M2 giant which rotates
along the M-theory $\phi$-circle. Without loss of generality, we can
choose $Z_4=e^{i\phi(t)}, Z_1=Z_2=Z_3=0$. The analysis is almost the
same as the previous case, but there are slight differences. The
energy is $E=kL/2$ and the size/angular momentum relation becomes
\begin{equation}\label{thingy}
\sinh\mu=\sqrt{\lambda\over 2}{kL\over N}\ ,
\end{equation}
where the angular momentum is $L\equiv {\partial L_{\rm
    M2}\over\partial\omega}$ with $\omega={d\phi\over dt}$. Upon
dimensional reduction to type IIA, this M2 giant becomes a bound state
of a D2-brane and $L$ D0-branes. Since the D0-branes are monopoles in
the dual field theory, the dual operator carries $L$ units of monopole
charge. The monopole operators can be labeled by the Cartan generators
$H={\rm diag} (q_1, q_2 \cdots, q_N)$ and $H'={\rm diag} (q'_1, q'_2
\cdots, q'_N)$ of two $U(N)$'s with $q_1\ge q_2\ge\cdots\ge q_N$ and
$q'_1\ge q'_2\ge\cdots\ge q'_N$. In particular, the monopole operator
${\cal M}_k$ with the unit charges $q_1=q'_1=1$ is in the
$k$-dimensional symmetric representations of two $U(N)$'s
\cite{Klebanov:2009sg}. Note that, when the level $k=1$, the operator
${\cal M}_1$ is bi-fundamental. Thus the dual operator is the Schur
polynomial of the $U(N)$ adjoint field $W_4\bar{{\cal M}}_1$ in the
$L$-dimensional symmetric representation. For $k>1$, the dual operator
is the gauge invariant constructed from $(W_4)^{kL}(\bar{{\cal
    M}}_k)^L$. This operator has dimension $kL/2$ which agrees with
the energy of the AdS giant.

%%%%%%%%%%%%%%%%%%%%%%%%%%%%%%%%%%%%%%%%%%%%%%%
\subsection{Holographic three-point functions}

The computation of the holographic three-point function between two
symmetric giants and a chiral primary operator proceeds similarly to
section \ref{antiholo}. We parametrize the $S^2$ in $AdS_4$ using (see
(\ref{poin})) $\vec n =
(\cos\vt,\sin\vt\sin\vp,\sin\vt\cos\vp)$. Using (\ref{fluct}) we find
that the variation of the Lagrangian density is
\bsp
\delta {\cal L}_{\rm DBI+WZ} = &\frac{T_{\rm M2}R^3_{\rm AdS}}{2}
\sinh\m \sin\vt\Biggl[-\frac{J}{3} s +
  h_{tt} + h_{\vt\vt} +
  \frac{h_{\vp\vp}}{\sin^2\vt}\Biggr]\\
&-2T_{\rm M2}R^3_{\rm AdS}\cosh\m\sinh^2\m \sin\vt \,\partial_\m s,
\end{split}
\ee
where $s= s^J(X)Y_J(\Omega)$ and where $\m$ is the global $AdS_4$
coordinate from (\ref{symmS}).

Using the chiral primary corresponding to (\ref{point}), i.e. the
point-like degeneration of the giant itself, we find the following
structure constant defining the extremal three-point function
\bsp\label{Csymm}
C_{L, L -\Delta, \Delta}^S = &\frac{1}{N}\left(\frac{\lambda}{2\pi^2}\right)^{1/4}
  2L\,\sqrt{2\Delta+1}  
 \\
& \times \left(1+\frac{2 L^2}{Nk}\right)^{-1-\Delta/2} 
{}_2F_1 \left( 1,1+\Delta,3/2,\frac{2L^2}{Nk+2L^2}\right). 
\end{split}
\ee
We also note the results for the structure constants
corresponding to the three-point functions of two symmetric giants and
one of the operators in (\ref{opss}), i.e. non-extremal correlators
\bsp\label{slo}
C^S_{{\cal O}_{\Delta,0}} = &\frac{1}{N}\left(\frac{\lambda}{2\pi^2}\right)^{1/4}
  2L\,\sqrt{2\Delta+1}\,
\sqrt{\pi} \frac{\Gamma(1+\Delta/2)}{\Gamma(1/2+\Delta/2)}  \, Y_{2\Delta}
 \\
& \times \left(1+\frac{2 L^2}{Nk}\right)^{-1-\Delta/2} 
{}_2F_1 \left( 1+\Delta/2,1+\Delta/2,3/2,\frac{2L^2}{Nk+2L^2}\right),  
\end{split}
\ee
where
$Y_{2\Delta}=-1/(2\sqrt{3}),\,-1/(12\sqrt{10}),\,3/(16\sqrt{35})$ for
$\Delta=1,2,3$ respectively. We note that the expressions
(\ref{Csymm}) and (\ref{Canti}) (and similarly, (\ref{slo}) and
(\ref{aslo})) agree in the point-like limit, when $L/N$ is small
(i.e. $\alpha \to 1/\sqrt{2}$). Thus the symmetric case also reduces
to the point-like result (\ref{abjmcpo}), in the large $J_2=J_3=2L$
limit. Comparing with free-field contractions (\ref{exttree}) and
(\ref{nonxgg})-(\ref{nonxgg1}), we see that, as in the point-like
case, a non-trivial function of $\lambda$ and the charges interpolates
between weak and strong coupling results.

%%%%%%%%%%%%%%%%%%%%%%%%%%%%%%%%%%%%%%%%%%%%%%%%%%%%%
\section{Conclusion \label{conclusion}}

Our greatly simplified parametrization of the anti-symmetric giant
graviton of $AdS_4\times \Cset\mathrm{P}^3$ made possible the
calculation of holographic three-point functions. It is possible that
further analytical results can now be obtained. One interesting
example is the possibility of obtaining an instanton solution
describing the tunneling of the anti-symmetric giant graviton to a
point-like one. Such a solution is known to exist in the $AdS_5\times
S^5$ background~\cite{Hashimoto:2000zp}.

Our holographic three-point functions involving two giant and one
point-like graviton reduce to the three-point function of three
point-like gravitons calculated in the supergravity approach when the
size of the giants approaches zero but remains larger than ${\cal
  O}(1)$. The supergravity three-point functions behave as
$\lambda^{1/4}$ as $\lambda\rightarrow \infty$ (where
$\lambda=\frac{N}{k}$) signaling that three-point functions of chiral
primaries in ABJM theory are not protected. Hence, we do not expect
to be able to recover our holographic three-point functions by a gauge
theory computation. In the case of ${\cal N}=4$ SYM, where
three-point functions of chiral primaries are known to be protected,
the method developed for calculating holographic three-point functions
of giant and point-like gravitons~\cite{Bissi:2011dc} led to a
complete match between gauge and string theory for non-extremal
correlators~\cite{Caputa:2012yj}, but extremal correlators did not
match their gauge theory duals completely~\cite{Bissi:2011dc}. Based
on these observations we expect that our non-extremal three-point
correlators correctly encode the strong coupling behaviour of ABJM
theory. It remains, however, of utmost importance to fully understand
the subtleties of holographic three-point functions in the extremal
case and we hope that our results for the $AdS_4\times\Cset
\mathrm{P}^3$ case will provide useful data for the future development
of this topic.

An interesting outcome of our analysis is that holographic three-point
functions are very different for anti-symmetric and symmetric giant
gravitons in $AdS_4\times \Cset\mathrm{P}^3$. This difference is not
reflected by the dual correlation functions in ABJM theory when
calculated at tree-level, cf.~appendix \ref{appendix}. As pointed out
above, in ABJM theory three-point functions of 1/2 BPS operators are
not protected and not even the lowest order loop correction to the
three-point function of chiral primaries is known. Calculating such
loop corrections constitutes another important future task.

\section*{Acknowledgments}

We thank A.\ Bissi, P.\ Caputa, H.\ Shimada and K.\ Zoubos for useful
discussions. SH would like to thank The Niels Bohr Institute for their
great hospitality. CK and DY were supported by FNU through grant
number 272-08-0329. SH was partially supported by the Grant-in-Aid for
Nagoya University Global COE Program (G07).

\appendix

%%%%%%%%%%%%%%%%%%%%%%%%%%%%%%%%%%%%%%%%%%%%%%%%%%%%%%%%%%%%%%%%%%%%%%
\section{Chiral primary structure constants from supergravity}
\label{app:cpo}

The strong coupling result for the three-point function structure
constant was given for the $k=1$ case in
\cite{Bastianelli:1999en}. Restoring the $k$ dependence is trivial, and
one obtains the following expression
\bsp
C_{123}^{\l\gg1}=
\frac{1}{N}\left(\frac{\l}{2\pi^2}\right)^{1/4} \frac{1}{\G(\g/2+1)}\,
\prod_{i=1}^3&
\frac{\G(\g_i/2+1)\,\sqrt{J_i+1}}{\sqrt{J_i!}}\\
&\times\frac{k\,2^\g\,(\g+3)!}{2^8\pi^4} 
\int_{S^7/\bZ_k} {\cal Y}_{J_1} {\cal Y}_{J_2} {\cal Y}_{J_3},
\end{split}
\ee 
where the $S^7/\bZ_k$ is taken to have radius 2, and the spherical harmonics $ {\cal Y}_{J_i}$
are taken to be normalized as 
\be
\int_{S^7/\bZ_k} {\cal Y}_J \bar{\cal Y}_K = \frac{2^8\pi^4}{k}\d_{JK} \frac{J!}{2^J(J+3)!}.
\ee
We would like to evaluate the integral of three spherical
harmonics, written using the ${\cal C}$ tensors appearing in the
definition of the operators (\ref{defop}), i.e. using the harmonics
\be
Y_J = ({\cal C}_{A})^{I_1\ldots I_{J/2}}_{K_1\ldots K_{J/2}} \,
Z_{I_1}\cdots Z_{I_{J/2}}  \bar Z^{K_1} \cdots\bar Z^{K_{J/2}}.
\ee
We use the following identity proven in \cite{Drukker:2008jm}
\be\label{sphid}
\int_{S^7/\bZ_k} Z_{I_1}\cdots Z_{I_m}\bar Z^{K_1}\cdots\bar Z^{K_m}
=\frac{2^8\pi^4}{k(m+3)!}\sum_{\sigma\in S_m}
\delta^{I_1}_{K_{\sigma(1)}}\cdots\delta^{I_m}_{K_{\sigma(m)}}\,,
\ee
to show that the $Y_J$ are normalized according to footnote
\ref{sphnrm}. The relation between the ${\cal Y}_J$ and the $Y_J$ is
then
\be
{\cal Y}_J = \sqrt{\frac{J!}{2^J}}\frac{1}{(J/2)!} \,Y_J.
\ee
We will also require the integral over three $Y_{J_i}$. The identity
(\ref{sphid}) instructs us to count all possible contractions between
the $Z$'s and $\bar Z$'s. We use the following
figure
\begin{center}
\includegraphics[bb=0 0 230 230,height=2in]{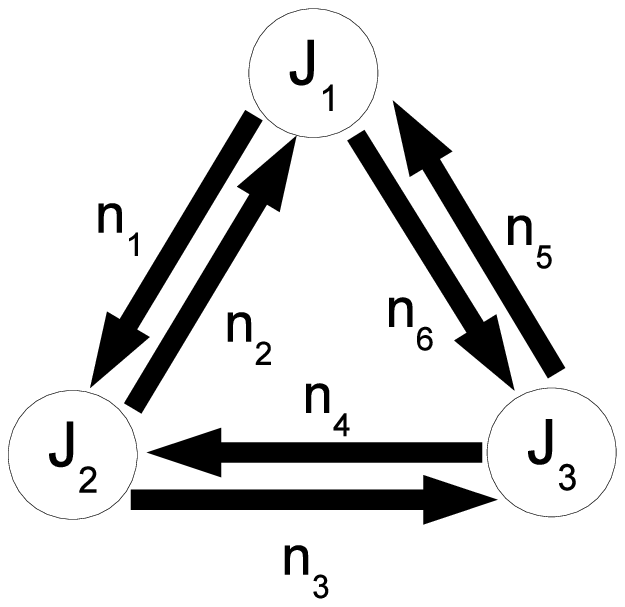}
\end{center}
to explain this counting. We have $J_i/2$ upper indices and $J_i/2$
lower indices in each of three ${\cal C}$ tensors defining the three
spherical harmonics. We denote upper-to-lower ($Z$ to $\bar Z$)
contractions with an arrow pointing at the lower index. We thus have
the following constraints
\bsp
&n_1 + n_6 = J_1/2 = n_2 + n_5,\\
&n_2+n_3 = J_2/2 = n_1+n_4,\\
&n_4 + n_5 = J_3/2 = n_3 + n_6,
\end{split}
\ee
which yields the solution
\bsp
&n_1 = p,\qquad
n_2 = \g_3 - p,\qquad
n_3 = \g_1 - \frac{J_2}{2} + p,\\
&n_4 =\frac{J_2}{2}-p,\qquad
n_5 = \g_2 - \frac{J_1}{2}+p,\qquad
n_6 = \frac{J_1}{2} - p,
\end{split}
\ee
where $\g_i = (\sum_j J_j -2J_i)/2$ denotes the total number of
contractions between the two operators other than the $i^\text{th}$.
Assuming, w.l.o.g. that $J_3\geq J_2\geq J_1$, we see that 
\be
p \in [0,\g_3].
\ee
We can now count possible contractions. We have $(J_1/2)!/[n_1!n_6!]$
ways of dividing the $J_1/2$ $Z$'s into two groups of $n_1$ and $n_6$
respectively, while for the $J_1/2$ $\bar Z$'s we have
$(J_1/2)!/[n_2!n_5!]$. Multiplying by similar factors for each of the
three operators, we have
\be
\left(\frac{\prod_{i=1}^3 (J_i/2)!}{\prod_{j=1}^6 n_j!}\right)^2
\ee 
many ways of splitting the various $Z$'s and $\bar Z$'s into their
requisite groups. We then have $\prod_{j=1}^6 n_j!$ ways of
contracting the groups together. We therefore find that
\bsp
&\int_{S^7/\bZ_k} Y_{J_1} Y_{J_2} Y_{J_3} = \frac{2^8\pi^4}{k\,(\g+3)!}
\sum_{p=0}^{\g_3}\\
&\frac{(J_1/2)!^2 (J_2/2)!^2 (J_3/2)!^2}{p! (\g_3-p)! (\g_1-J_2/2+p)!
(J_2/2-p)!(\g_2-J_1/2+p)! (J_1/2-p)!}\\
&\left({\cal C}_{J_1}\right)^{I_1\ldots I_p I_{p+1}\ldots I_{J_1/2}}
_{K_1 \ldots K_{\g_3 -p} K_{\g_3-p+1} \ldots K_{J_1/2}}\\
&\left({\cal C}_{J_2}\right)^{K_1 \ldots K_{\g_3 -p} L_1\ldots
L_{\g_1-J_2/2+p}}_{I_1\ldots I_p M_1 \ldots M_{J_2/2-p}}
\left({\cal C}_{J_3}\right)^{K_{\g_3-p+1} \ldots K_{J_1/2}M_1 \ldots M_{J_2/2-p}}
_{I_{p+1}\ldots I_{J_1/2}  L_1\ldots L_{\g_1-J_2/2+p}},
\end{split}
\ee
and the expression (\ref{abjmcpo}) follows.

We note that there are also other spherical harmonics with an unequal
number $\D^+$ of $Z$'s and $\D^-$ of $\bar Z$'s such that $\D^+ -
\D^-= mk$ where $m$ is an integer. These correspond to states in ABJM
with non-zero $U(1)_B$ charge, discussed for example in
\cite{Drukker:2008jm}, and which require the presence of the monopole
operators discussed beneath (\ref{thingy}). Our analysis can also be
carried out for this more general case using $n_1+n_6 = \D^+_1$,
$n_2+n_5=\D^-_1$, etc., and requiring that the total $U(1)_B$ charge,
$\sum_i m_i=0$. This then gives a generalization of (\ref{abjmcpo})
for these more general operators. It is not clear however, how to
evaluate these more general structure constants in perturbation theory.

%%%%%%%%%%%%%%%%%%%%%%%%%%%%%%%%%%%%%%%%%%%%%%%%%%%%%%%%%%%%%%%%%%%%%%
\section{The dual operators and their three-point functions
\label{appendix}}
In ABJM theory one can construct operators which form Schur polynomials
of a single $U(N)$ by 
combining two bi-fundamental scalar fields. Denoting 
the two complex bi-fundamental scalars as $W_1$ and $\bar W^2$ a $U(N)$
Schur polynomial can then be written as
\begin{equation}
\chi_{R_L}(W_1\bar W^2)=\frac{1}{L!}
\sum_{\sigma\in S_L} \chi_{R_L}(\sigma) \,(W_1\bar{W}^2)_{i_1}^{i_{\sigma(1)}}
\ldots (W_1\bar{W}^2)_{i_L}^{i_{\sigma(L)}},
\label{Schur}
\end{equation}
where $R_L$ denotes an irreducible representation of $U(N)$ described in terms
of a Young tableau with $L$ boxes. The sum is over elements of the 
symmetric group and $\chi_{R_L}(\sigma)$ is the character of the element 
$\sigma$ in the representation $R_L$. The calculation of two- and three-point
functions of operators of the type~(\ref{Schur}) at tree-level is a purely
combinatorial problem which can be 
solved in close analogy with the similar
problem involving a single adjoint scalar appearing in ${\cal N}=4$ 
SYM, see~\cite{Dey:2011ea}.

The structure constant dual to the three-point function of two 
giants
and one point-like graviton of $AdS_4\times \Cset\mathrm{P}_3$ is
\begin{equation}
C_{L,L-\Delta,\Delta} \equiv
\frac{\langle \bar{\chi}_L \,
\chi_{L-\Delta} \Tr (W_1\bar{W}^2)^{\Delta} \rangle}
{\sqrt{\langle \bar{\chi}_L\, \chi_{L} \,\rangle 
\langle \bar{\chi}_{L-\Delta} \chi_{L-\Delta}\rangle
\langle \Tr (\bar W^1W_2)^{\Delta} \Tr (W_1\bar{W}^2)^{\Delta}\rangle
}},
\end{equation}
where here and in the following
the expectation values are to be understood as expectation values
in a zero-dimensional Gaussian complex matrix model with unit propagator.
Furthermore, $\chi_{L}$ is the Schur polynomial 
corresponding to a Young tableau consisting either of a
single column (anti-symmetric case) or a single row (symmetric case) 
with $L$ boxes 
and we have suppressed the dependence of the $W_I$-fields.
 Expanding the
single trace operator in the basis of Schur polynomials and making use of the
known three-point functions of the Schurs from~\cite{Dey:2011ea} one easily finds
the following expression for the three-point functions in the limit 
$\Delta\ll L$, $L, N\rightarrow \infty$, $\frac{L}{N}$ fixed
\begin{equation}\label{exttree}
C_{L,L-\Delta,\Delta}^A=\frac{1}{\sqrt{\Delta}}
\left(1-\frac{L}{N}\right)^{\Delta},
\hspace*{0,7cm}
C_{L,L-\Delta,\Delta}^S=(-1)^{\Delta-1}\frac{1}{\sqrt{\Delta}}
\left(1+\frac{L}{N}\right)^{\Delta},
\end{equation}
where the superscript 
$A$ refers to the anti-symmetric case and $S$ to the symmetric one.
For details we refer to~\cite{Bissi:2011dc}.

To determine the tree-level contribution to the non-extremal
ABJM three-point function, dual to the 
correlator of two giant gravitons and one point-like one of the type
given in equation~(\ref{opss})  one has to evaluate
\begin{equation}
C_{{\cal O}_{\Delta,0}}=
\frac{\langle \bar{\chi}_L \,
\chi_{L} \,{\cal O}_{\Delta,0} \rangle}
{\langle \bar{\chi}_L\, \chi_{L} \,\rangle
}.\label{CDelta}
\end{equation}
Writing the operators out
explicitly, stripping off the factors originating from the gauge theory
propagators
 and furthermore exploiting the symmetry properties of the 
expectation values one finds that in the formula~(\ref{CDelta}) one can 
replace  ${\cal O}_{\Delta,0}$ by ${\cal O}_{\Delta,0}^{eff}$ given by
\begin{eqnarray}
{\cal O}_{1,0}^{eff}&=& \frac{1}{2\sqrt{3} N} 
 \Tr (-2W_1\bar W^1),\\
{\cal O}_{2,0}^{eff}&=& \frac{1}{6\sqrt{5}N^2}
\Tr \left[ 7(W_1\bar W^1)^2
-4W_1\bar W^1 W_2\bar W^2 -4W_1\bar{W}^2 W_2\bar W^1\right],\\
{\cal O}_{3,0}^{eff}&=& \frac{1}{12\sqrt{105}\,N^3}
\Tr \left[- 9(W_1\bar W^1)^3+5(W_1\bar W^1)^2 W_2\bar W^2 
\right.\nonumber\\
&&
\left.
+5W_1\bar{W^1}W_1\bar{W}^2 W_2\bar W^1
+5W_1\bar{W^1}W_2\bar W^1 W_1\bar{W}^2
\right].
\end{eqnarray}
A somewhat lengthy but in principle straightforward
calculation along the lines 
of~\cite{Caputa:2012yj} gives
\begin{eqnarray}\label{nonxgg}
C_{{\cal O}_{1,0}}^S&=&C_{{\cal O}_{1,0}}^A= 
-\frac{1}{\sqrt{3}}\frac{L}{N}, \\
C_{{\cal O}_{2,0}}^S&=& -\frac{1}{6\sqrt{5}}\frac{L}{N}
\left(8+\frac{L}{N}\right),\hspace*{0.7cm}
C_{{\cal O}_{2,0}}^A= 
-\frac{1}{6\sqrt{5}}\frac{L}{N}
\left(8-\frac{L}{N}\right), \\
C_{{\cal O}_{3,0}}^S&=& \frac{1}{12\sqrt{105}}\,\frac{L}{N}
\left(6\left(\frac{L}{N}\right)^2+5+20\frac{L}{N} \right),\\
C_{{\cal O}_{3,0}}^A&=& \frac{1}{12\sqrt{105}}\,\frac{L}{N}
\left(6\left(\frac{L}{N}\right)^2+5-20\frac{L}{N} \right)\label{nonxgg1}.
\end{eqnarray}

\bibliography{ABJMgiant}

\end{document}